\documentclass[pre,twocolumn,showpacs,showkeys,preprintnumbers,amsmath,amssymb]{revtex4-1}
\usepackage{graphicx}
\usepackage{float}
\usepackage{bm}

\begin{document}

\title{Monte Carlo simulations of dynamic phase transitions in ferromagnetic thin-films}

\author{Bahad\i r~Ozan~Akta\c{s}}
\email{ozan.aktas@ogr.deu.edu.tr} \affiliation{Dokuz Eylul
University, Graduate School of Natural and Applied Sciences,
TR-35160 Izmir, Turkey}
\author{Erol Vatansever}
\author{Hamza Polat}
\affiliation{Department of Physics, Dokuz Eyl\"{u}l University,
TR-35160 Izmir, Turkey}

\date{\today}

\begin{abstract}
By means of detailed Monte Carlo (MC) simulations, we have presented
dynamic phase transition (DPT) properties of ferromagnetic
thin-films. Thermal variations of surface, bulk and total dynamical
order parameters (DOP) for a film and total order parameter for the
films with different thicknesses have been examined. Opposite
regimes of the critical value of reduced exchange interaction
(surface to bulk ratio) $R_{c}$ at which the critical temperature
becomes independent of film thickness $L$ has been also taken into
consideration. The average magnetizations of each layer is reversed
in these regimes. Based on the results, we have confirmed that the
system represents a crossover behavior in between ordinary to
extraordinary transition in the presence of surface exchange
enhancement.
\end{abstract}


\maketitle

\section{Introduction}\label{intro}

Magnetic properties of free surfaces drastically differ from the
bulk material, because the free surface breaks the translational
symmetry (i.e. surface atoms are embedded in an environment of lower
symmetry than that of the inner atoms and consequently the exchange
constants between atoms in the surface region may differ from the
bulk value). The surface enhancement phenomenon in finite magnetic
materials has attracted considerable amount of interest for both
experimentalists \cite{schierle, ahlberg1, violbarbosa, berger,
wang, yalcin} and theorists \cite{saber1, oubelkacem1, tuleja, zaim,
akinci1, yuksel1, akinci2, akinci3, yuksel2, yuksel3}.

Applied oscillating magnetic fields, depending on the competition
between the two time scales, namely the oscillation period $P$ of
the external perturbation and the relaxation time $\tau$ of the
sample, a dynamic symmetry breaking may take place causing a DPT.
There are two cases due to the competition between these time
scales: $P<\tau$ and $P>\tau$. In the first case, the system cannot
relax within a complete cycle of the magnetic field oscillation,
hence the instantaneous magnetization $M(t)$ oscillates in time
around a nonzero value corresponding to dynamically ordered state
(i.e. dynamic ferromagnetic phase). In other case, $M(t)$ can follow
the external field with some delay, and the system exhibits a
dynamic paramagnetic behavior. The relaxation time $\tau$ can be
controlled by supplied energy with several different ways: The
agency of an adjustable parameter such as the field amplitude,
strength itself, the type of the exchange interactions, and the
temperature. The DPT point can be controlled by tuning mentioned
competing factors together with the time period of external field.

Experimental point of view, Schierle and coworkers observed that the
magnetizations of the outermost layers in EuTe(111) films decrease
significantly differently from those of bulk layers \cite{schierle}.
Violbarbosa and coworkers found that the formation of the blocks of
layers with robust magnetic structure whereas the interblock
interactions are relatively weak in fcc-Fe on Cu(001) film
\cite{violbarbosa}. Moreover, the enhanced surface magnetism has
been the focus of such systems. For instance, Gd film has been
investigated experimentally. The thickness-dependent spin-polarized
electronic structure of strained ultrathin and thin films of Gd has
been investigated by Waldfried et al. \cite{waldfried}. They found
that the surface magnetic structure dominates the magnetic ordering
of the ultrathin Gd films. With decreasing thickness some bulk bands
exhibit increasingly more passive magnetic behavior. Skomski and
coworkers also found that Gd films exhibit a magnetic surface
transition which occurs at about above the bulk Curie temperature
\cite{skomski}.

In sense of dynamic phase transitions, a great deal of theoretical
efforts has also been devoted to the investigation on such systems.
The details of surface enhancement phenomenon for the films were
subjected to an external oscillatory field have been intensively
propounded by Akta\c{s} et al. by using effective field theory (EFT)
\cite{aktas1}. The general trend of frequency dispersion belongs to
critical temperature coordinate of the special point for different
frequency and amplitude values has been demonstrated in their work.
Nonequilibrium phase transition in the kinetic Ising model on a
two-layer square lattice has been examined by Canko et al.
\cite{canko}. Dynamic phase diagrams have been constructed in the
plane of the reduced temperature versus the amplitude. Similarly,
dynamic magnetic behavior of a mixed Ising system on a bilayer
square lattice has been investigated by Erta\c{s} and Keskin
\cite{ertas}. They presented the dynamic phase diagrams in the
reduced temperature and magnetic field amplitude plane and the
effects of interlayer coupling interaction on the critical behavior
of the system have been investigated in their work.

In recent series of works by Pleimling and coworkers, surface
criticality at a DPT and surface phase diagram of the
three-dimensional kinetic Ising model has been elucidated. In the
first one of these studies, Park and Pleimling found that the
nonequilibrium surface exponents do not coincide with those of the
equilibrium critical surface \cite{park1}. In addition, in three
space dimensions, the surface phase diagram of the nonequilibrium
system differs markedly from that of the equilibrium system. The
values of the critical exponents have been determined through
finite-size scaling by Park and Pleimling in their followup
investigation \cite{park2}. Their results have showed that the
studied nonequilibrium phase transition belongs to the universality
class of the equilibrium three-dimensional Ising model. The surface
phase diagram of the three-dimensional kinetic Ising model below the
equilibrium critical point subjected to a periodically oscillating
magnetic field has also been presented by Taucher and Pleimling
\cite{taucher}. They presented that surface phase diagram that in
parts strongly resembles the corresponding equilibrium phase
diagram, with an ordinary transition, an extraordinary transition,
and a surface transition. These three lines meet at a special
transition point. For weak surface couplings, however, the surface
does not order.

In this regard, our task in the present work is to shed some light
on the DPT properties, -especially the evolution of crossover point
with field amplitude- of ferromagnetic thin-films in the presence of
ac driving fields. In this present paper, the DPT properties in the
presence of external oscillatory field of the system are studied by
MC simulation. The layout of the work is as follows: Section 2
describes the model and the MC simulation scheme, the numerical
results are reported in Section 3, the paper ends with concluding
remarks in Section 4.

\section{Simulation}\label{method}
We consider a ferromagnetic thin film with thickness $L$ described
by spin-1/2 Hamiltonian
\begin{equation}\label{eq1}
\mathcal{H}=-\sum_{\langle ij\rangle}J_{ij}s_{i}s_{j}-h(t)\sum_{i} s_{i}
\end{equation}
where $s_{i}=\pm1$ is a two-state spin variable, and $J_{ij}$ is the
nearest neighbor interaction energy. The summation in the first term
is taken over only the nearest neighbor interactions whereas the
summation in the second term is carried out over the all lattice
sites. In the second term, $h(t)=h_{0}\sin(\omega t)$ represents the
oscillating magnetic field, where $h_{0}$ and $\omega$ are the
amplitude and the angular frequency of the applied field,
respectively. The period of the oscillating magnetic field is given
by $P=2\pi/\omega$. If the lattice sites $i$ and $j$ belong to one
of the two surfaces of the film we have $J_{ij}=J_{s}$, otherwise
$J_{ij}=J_{b}$, where $J_{s}$ and $J_{b}$ denote the ferromagnetic
surface and bulk exchange interactions, respectively.

In order to simulate the system, we employ the Metropolis MC
simulation algorithm \cite{binder1, newman1} to Eq. (\ref{eq1}) on
an $N\times N\times L$ simple cubic lattice where $N=70$ and we
apply periodic (free) boundary conditions in direction(s) parallel
(perpendicular) to film plane. We have studied ultrathin-films with
thickness $L=3,4,5$ together with a relatively thicker thin-film
$L=20$ to observe average magnetizations of each layer for selected
some system parameters.  For simplicity, the exchange couplings are
restricted to the ferromagnetic case.

Configurations were generated by selecting the sites in sequence
through the lattice and making single-spin-flip attempts, which were
accepted or rejected according to the Metropolis algorithm, and
$N\times N\times L$ sites are visited at each time step (a time step
is defined as an MC step per site or simply MCS). Data were
generated over 50 independent sample realizations by running the
simulations for 50000 MCS per site after discarding the first 25000
steps.  This amount of transient steps is found to be sufficient for
thermalization for the whole range of the parameter sets. Throughout
the analysis, oscillation period of the external field is kept fixed
as $P=100$.

Our program calculates the instantaneous values of the bulk and
surface magnetizations $M_{s}$ and $M_{b}$, and the total
magnetization $M_{T}$ at time $t$. These quantities are defined as
\begin{equation}\label{eq2}
\begin{split}
M_{s}(t)=\frac{1}{N_{s}}\sum_{i=1}^{N_{s}}s_{i},\quad
M_{b}(t)=\frac{1}{N_{b}}\sum_{j=1}^{N_{b}}s_{j},\\
M_{T}(t)=\frac{N_{s}M_{s}(t)+N_{b}M_{b}(t)}{N_{s}+N_{b}}\quad\quad\quad
\end{split}
\end{equation}
where $N_{s}$ and $N_{b}$ denote the number of spins in the surface
and bulk layers, respectively. From the instantaneous
magnetizations, we obtain the related order parameters as follows
\cite{tome}:
\begin{equation}\label{eq2}
\begin{split}
Q_{s}=\frac{1}{P}\oint M_{s}(t)dt,\quad
Q_{b}=\frac{1}{P}\oint M_{b}(t)dt,\\
Q_{T}=\frac{1}{P}\oint M_{T}(t)dt\quad\quad\quad\quad\quad
\end{split}
\end{equation}
Using Eq. (\ref{eq1}), we calculate the total energy per spin
\begin{equation}\label{eq3}
E_{tot}=\frac{1}{P(N_{s}+N_{b})}\oint \mathcal{H} dt
\end{equation}
Consequently, the specific heat is defined as
\begin{equation}\label{eq4}
C=\frac{dE_{tot}}{dT}.
\end{equation}
We also note that the value of the bulk exchange interaction $J_{b}$
is fixed to unity, and we also use the normalized surface to bulk
ratio of exchange interactions $R=J_{s}/J_{b}$, as well as the
reduced field amplitude $H_{0}=h_{0}/J_{b}$, and reduced temperature
$\Theta=k_{B}T/J_{b}$.

\section{Results and Discussion}\label{result}
\begin{figure}[H]
\centering
\includegraphics[width=8cm]{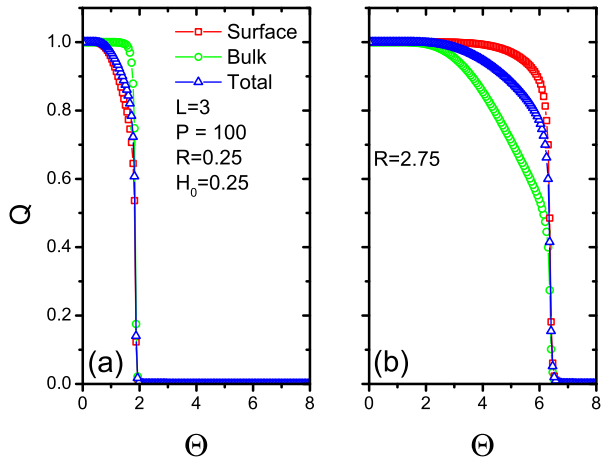}
\caption{(Color online) Surface, bulk and total order parameters for
a film $L=3$ layers for two different regime of reduced exchange
interaction.}\label{fig1}
\end{figure}

\begin{figure}[H]
\centering
\includegraphics[width=8.5cm]{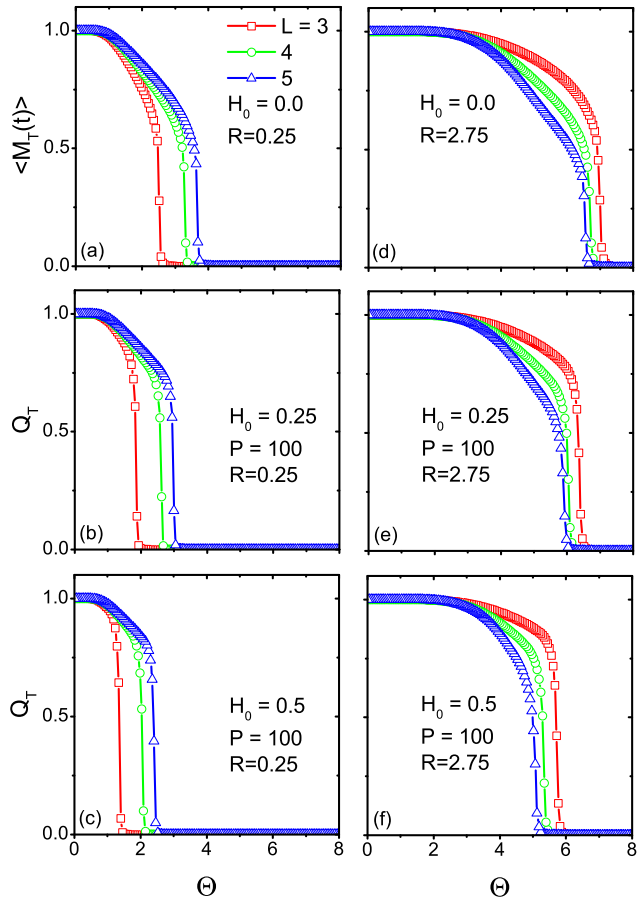}
\caption{(Color online) Average magnetization in static case (a) and
(d), and the total order parameters for the films with three
different thicknesses $L=3,4,5$. and for three different values of
$H_{0}$.}\label{fig2}
\end{figure}

Based on the results, we mainly focus on the effect of external
oscillatory field amplitude in surface enhancement phenomenon.
First, in order to obtain a general insight on DPT characteristics,
we plot Fig. (\ref{fig1}) and (\ref{fig2}) respectively. In Fig.
(\ref{fig1}) the thermal variations of surface, bulk and total order
parameters for the films with thickness $L=3$ are shown. Following
this, total dynamical order parameters for the films with three
different thicknesses are shown in Fig. (\ref{fig2}). We restrict
our discussions for two value of reduced exchange $R=0.25$ and
$2.75$. Both in Fig. (\ref{fig1}) and (\ref{fig2}), the transition
point increases with reduced exchange interaction at constant values
of the other system parameters. Hence, relatively more thermal
agitation is needed to make the system dynamically disordered for
more interaction. Moreover, at a constant temperature, surfaces are
weakly ordered due to the scarcity of dipole-dipole interaction per
site for $R=0.25$ value. For $R=2.75$, surfaces dominate against
bulk due to the surface enhancement. From Fig. (\ref{fig1}), we see
that both surfaces and bulk layers of magnetic thin-films exhibit a
phase transition at a certain critical temperature independently
from the value of $R$.

From Fig. (\ref{fig2}), one can easily see the effect of external
field amplitude $H_{0}$ for fixed value of the system parameters
from the panel (a) to (c) and (d) to (f). Critical temperature
exhibits a decreasing behavior with increasing $H_{0}$ as a
consequence of the well-known following physical mechanism: For
small amplitude values, the energy supplied by the external
oscillatory field cannot break the ferromagnetic energy induced
order due to the nearest-neighbor exchange coupling through the
system at low temperatures. Hence, a DPT cannot be observed unless a
relatively large amount of thermal energy is supplied to the system.
As the field amplitude increases, it becomes dominant against the
ferromagnetic nearest-neighbor bonds, and a DPT can be observed at
low temperatures. When we fixed rest of the parameters except
reduced exchange and compared the strength of the order parameters
for the films with different thicknesses in two different value of
$R$ (namely $R=0.25$ and $2.75$), we see the hierarchically sequence
reversal in Fig. (\ref{fig2}) (from (a) to (d) and (b) to (e) and
(c) to (f)). In between these two value, there should be a critical
point of the reduced exchange at which all the layers seem to
oscillate in phase independently from the thickness.

\begin{figure}
\centering
\includegraphics[width=8cm]{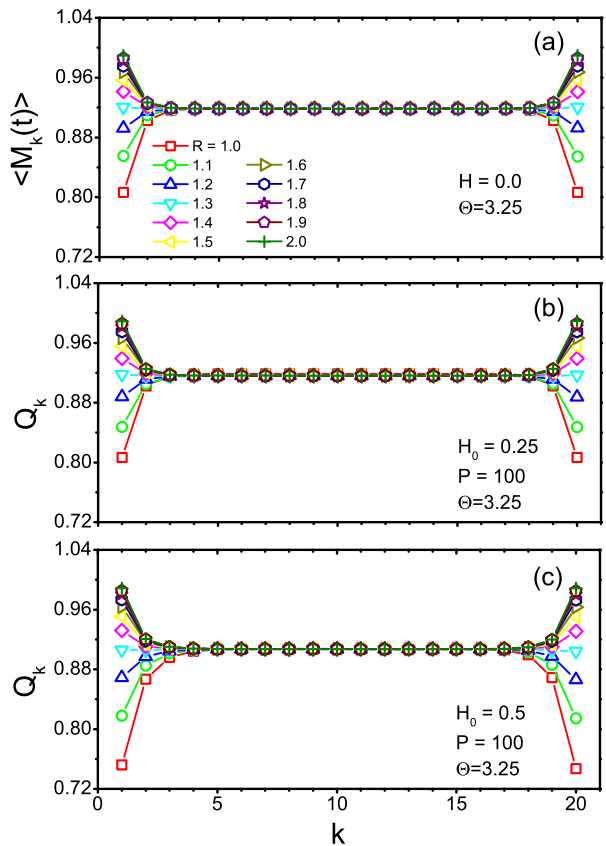}
\caption{(Color online) Average magnetizations profiles for a film
with $L=20$ layers. The number accompanying each curve denotes
several values of reduced exchange varies from 1.0 to 2.0 including
$R_{c}$. Each layer labelled by $k$.}\label{fig3}
\end{figure}

In order to make the aforementioned phenomenon more clear, we plot
the dynamical order parameters of each layers for a film with $L=20$
in Fig. (\ref{fig3}). For this purpose, we choose a constant
temperature value at which the system is well-below the transition
point and the thermal fluctuations can be ruled out. The
magnetization $M(t)$ cannot follow the external field $h(t)$
($\tau>P$ case) for each selected field amplitude values $H_{0}$,
consequently the dynamic ferromagnetism is enhanced. Reduced
exchange varies from $R=1.0$ to $2.0$ including the critical value
of itself. So, we have qualitatively different two regimes:
$R<R_{c}$ and $R>R_{c}$. Below $R_{c}$ the inner layers are highly
ordered compare to surfaces. This can be briefly explained as
follows: In middle of the film, there are relatively more
neighboring per magnetic sites which causes locally larger magnetic
interaction. So it becomes more difficult to follow the external
field for any spin. Surface spins are embedded in an environment of
lower symmetry than that of the inner atoms. The exchange constant
between atoms in the surface region may differ from the bulk one.
This regime corresponds to a surface type of magnetic ordering. The
opposite of the above scenarios can be considered also. Above
$R_{c}$, in the inner layers, although there are more neighboring,
there are far fewer exchange constant per magnetic sites which
causes relatively smaller magnetic interaction than that of the
surface one. $R_{c}$ plays the main role to obtain the frontier of
this crossover. The free surface cannot break the translational
symmetry since magnetic properties of the free surfaces exactly
overlap with the bulk one at $R_{c}$. We can say more generally that
the deficiency of the interaction per surface spin can be
compensated by increasing the modified exchange interaction
strength. Moreover, the effect of external field can be also seen by
following the panels from (a) to (c).

\begin{figure}
\centering
\includegraphics[width=8.5cm]{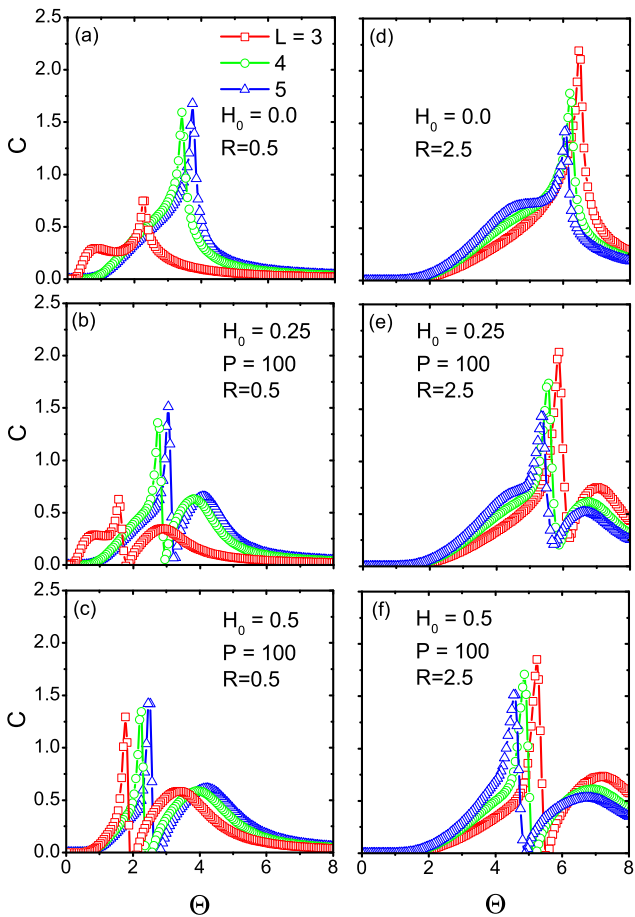}
\caption{(Color online) The specific heat versus temperature for the
films with different $L=3,4,5$ thicknesses.}\label{fig4}
\end{figure}

In obtaining the critical frontiers depicted in $(k_{B}T_{c}/J-R)$
plane, we evaluated the thermal variation of specific heat for a
given set of system parameters. A typical example is shown in Fig.
(\ref{fig4}) for the films with different thicknesses as $L=3,4,5$.
The temperature values corresponding to the maxima of specific heat
curves are the transition temperatures. From the panel (a) to (c)
and (d) to (f), field amplitude $H_{0}$ changes. Also, from (a) to
(d), (b) to (e) and (c) to (f) reduced exchange has two different
value. DPT points shift towards the lower temperature with
increasing field values as well as they shift towards higher
temperature with increasing reduced exchange. The related detailed
story has been explained above. Similarly, the effect of reduced
exchange on specific heat peaks is easy to understand: The stronger
dipole-dipole interaction make more contribution to total energy.
Hence, more thermal agitation is needed to make the system
dynamically disordered. In Fig. (\ref{fig4}), the crossover behavior
can be seen easily when $R$ changed from $R<R_{c}$ to $R>R_{c}$
(namely, from (a) to (d), (b) to (e) and (c) to (f)). Below $R_{c}$,
thicker film has more nearest-neighbor interaction per site, this
creates more contribution to energy. Consequently, both strength of
the peak and corresponding critical temperature are relatively
higher than the others.

\begin{figure}
\centering
\includegraphics[width=7.5cm]{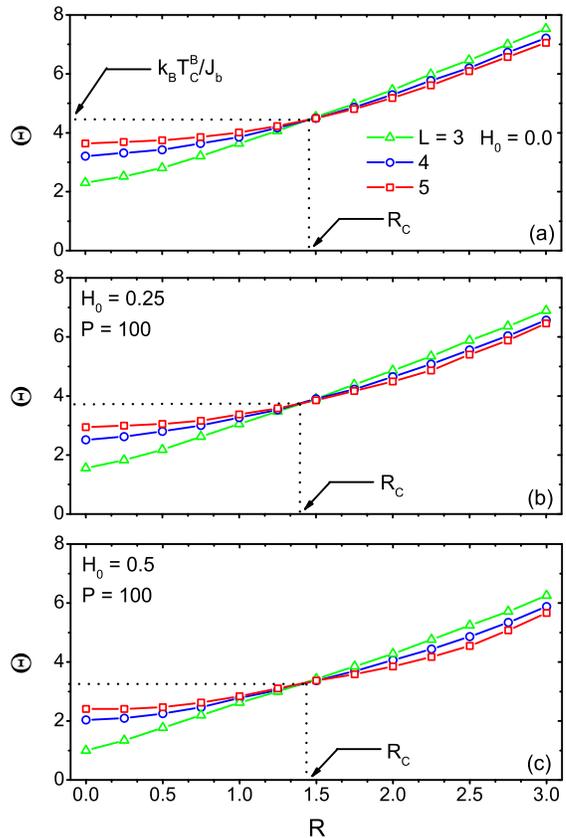}
\caption{(Color online) The critical temperature versus reduced
exchange interaction in phase planes for the films with different
$L=3,4,5$ thicknesses.}\label{fig5}
\end{figure}

In order to obtain a general overview of the non-equilibrium phase
diagram in $(k_{B}T_{c}/J_{b}-R)$ planes. For this purpose, in Fig.
(\ref{fig5}) we plot the critical temperature versus $R$ with
selected film thickness values $L=3,4,5$ and for three selected
values of field amplitude $H_{0}$. Since, the temperature values at
which the specific heat curves exhibit a sharp maximum correspond to
the transition temperature of thin film, critical temperature values
have been obtained by examining the thermal variation of specific
heat curves (a selected set has been given in Fig. (\ref{fig4})).
Fig. (\ref{fig5}) represents a characteristic phenomenon peculiar to
thin film systems. Namely, due to the existence of reduced surfaces,
there exists a special value of surface to bulk ratio of exchange
interactions $R{c}$ at which the transition temperature of the film
becomes independent of thickness $L$. Simply we can say that the
curves with different film thicknesses intersect each other. This
fully supports a recent study \cite{aktas1} where in the framework
of an EFT the existence of a special transition point was predicted.
The critical temperature value of crossover point for $H_{0}=0.0$
(static case) is in a good agreement with previous studies
\cite{kaneyoshi3, park1, aktas1}. However, variation of $R_{c}$ as a
function of $H_{0}$ is very slow according to Fig. (\ref{fig5}), and
we see that the location of $R_{c}$ barely deviates from its
equilibrium value with increasing $H_{0}$. This deviation has been
reported before by Y{\" u}ksel \cite{yuksel1}. The discussion on
existence of this kind of deviation is an academic issue and this
may be due to insufficient data and cannot be located accurately
with such an approach with less effort. This was also reported in a
MC simulation treatment of surface critical phenomena by Hasenbusch
\cite{hasenbusch}. From panel (a) to (c) the effect of field
amplitude $H_{0}$ on the transition characteristics of the film can
be seen and it is also straightforward as stated before: Greater the
amplitude $H_{0}$ means more energy transferred to the system over
half cycle by the external oscillatory field and this makes
transition from dynamically ordered to disordered phase more easy.
For $R<R_{c}$, we have ordinary transition behavior where the bulk
magnetism is dominant against the surface magnetism whereas for
$R>R_{c}$, the surface may exhibit enhanced magnetic behavior in
comparison with bulk. This is called extraordinary transition.
Moreover, as shown in Fig. (\ref{fig5}), for $R<R_{c}$, thicker
films have greater transition temperatures while for $R>R_{c}$, the
transition temperature of the film decreases with increasing
thickness. The results indicate that the well-known surface
enhancement properties of the system may change its characteristics
in the presence of an external oscillatory field.

\section{Conclusion}\label{conclude}

In conclusion, we have applied MC simulations to study the DPT
characteristics in thin ferromagnetic films in the presence of
oscillating magnetic fields. The foremost results obtained from
simulation data can be summarized as follows: We first investigate
the thermal variations of related order parameters for the films
with different thicknesses. The effect of the field amplitude and
reduced exchange on the previous standard arguments for static case
has been propounded. In the vicinity of the dynamic
ferromagnetic-paramagnetic phase transition temperature, specific
heat curves exhibit a sharp peak which becomes more apparent for
sufficiently high reduced exchange values ($R>R_{c}$ regime) values.
The thinner films in the absence of enhanced surfaces ($R<R_{c}$
regime) with high field amplitudes exhibit a weak peak in relatively
lower temperatures.

According to our findings, an increment of the field amplitude
causes a decreasing in corresponding temperature coordinate of the
crossover point in ($k_{B}T_{c}/J_{b}-R$) planes. Critical value of
surface to bulk ratio of exchange interactions $R_{c}$ at which the
transition temperature is independent of film thickness is not
apparently responsive to varying field amplitude values, but
exhibits slow variation as a function of $H_{0}$. We confirmed the
general trend was generated by using EFT calculations before
\cite{aktas1}. Hence, we can say that the evolution of a crossover
is not from the limitation of EFT.

We hope that this study will shed light on further investigations of
the dynamic nature of critical phenomena in pure crystalline
ferromagnetic thin films and will be beneficial from both
theoretical and experimental points of view.

\section*{Acknowledgements}
The numerical calculations in this paper were performed at
T\"{U}B\.{I}TAK ULAKB\.{I}M (Turkish agency), High Performance and
Grid Computing Center (TRUBA Resources) and this study was completed
at Dokuz Eylül University, Graduate School of Natural and Applied
Sciences. One of the authors (B.O.A.) would like to thank the
Turkish Educational Foundation (TEV) for full scholarship.


\begin{thebibliography}{99}
\bibitem{schierle} E. Schierle, E. Weschke, A. Gottberg, W. S{\" o}llinger, W. Heiss, G. Springholz, and G. Kaindl, Phys. Rev. Lett. 101, (2008) 267202.
\bibitem{ahlberg1} M. Ahlberg, M. Marcellini, A. Taroni, G. Andersson, M. Wolff, and B. Hj{\" o}rvarsson, Phys. Rev. B 81, (2010) 214429.
\bibitem{violbarbosa} C. E. ViolBarbosa, H. L. Meyerheim, E. Jal, J.-M. Tonnerre, M. Przybylski, L. M. Sandratskii, F. Yildiz, U. Staub, and J. Kirschner, Phys. Rev. B 85, (2012) 184414.
\bibitem{berger} A. Berger, O. Idigoras, and P. Vavassori, Phys. Rev. Lett. 111, (2013) 190602.
\bibitem{wang} B.-Y. Wang, J.-Y. Hong, K.-H. O. Yang, Y.-L. Chan, D.-H. Wei, H.-J. Lin, and M.-T. Lin, Phys. Rev. Lett. 110, (2013) 117203.
\bibitem{yalcin} O. Yal\c{c}\i n, \c{S}. {\" U}nl{\" u}er, S. Kazan, M. {\" O}zdemir, Y. {\" O}ner. JMMM 373 (2015) 144.
\bibitem{saber1} M. Saber, A. Ainane, F. Dujardin, B. Stébé, Journal of Non-Crystalline Solids 250 (1999) 735.
\bibitem{oubelkacem1} A. Oubelkacem, A. Ainanea, J. J. de Miguel, J. Ricardo de Sousa, M. Saber, Physica A 358 (2005) 160.
\bibitem{tuleja} S. Tuleja, J. Kecer, and V. Ilkoviç, Phys. Stat. Sol. (b) 243, (2006) 1352.
\bibitem{zaim} A. Zaim, M. Kerouad, Y. EL Amraoui, D. Baldomir, JMMM 316 (2007) e306.
\bibitem{akinci1} {\" U}. Ak\i nc\i, JMMM 329 (2013) 178.
\bibitem{yuksel1} Y. Y{\" u}ksel, Phys. Lett. A 377 (2013) 2494.
\bibitem{akinci2} {\" U}. Ak\i nc\i, JMMM 368 (2014) 36.
\bibitem{akinci3} {\" U}. Ak\i nc\i, Thin Solid Films 550 (2014) 602.
\bibitem{yuksel2} Y. Y{\" u}ksel, {\" U}. Ak\i nc\i, Physica B 433 (2014) 96.
\bibitem{yuksel3} Y. Y{\" u}ksel, Physica A 396 (2014) 9.
\bibitem{waldfried} C. Waldfried, T. McAvoy, D. Welipitiya, P. A. Dowben, E. Vescovo, Europhys. Lett. 42 (1998) 685.
\bibitem{skomski} R. Skomski, C. Waldfried, P. A. Dowbeny, J. Phys.: Condens. Matter 10 (1998) 5833.
\bibitem{aktas1} B. O. Akta\c{s}, {\" U}. Ak\i nc\i, H. Polat, Thin Solid Films 562 (2014) 680.
\bibitem{canko} O. Canko, E. Kantar, M. Keskin, Physica A 388 (2009) 28.
\bibitem{ertas} M. Erta\c{s}, M. Keskin, Chin. Phys. B 22 (2013) 120507.
\bibitem{park1} H. Park, M. Pleimling, Phys. Rev. Lett. 109, (2012) 175703.
\bibitem{park2} H. Park, M. Pleimling, Phys. Rev. E 87, (2013) 032145.
\bibitem{taucher} K. Taucher, M. Pleimling, Phys. Rev. E 89, (2014) 022121.
\bibitem{binder1} K. Binder, Monte Carlo Methods in Statistical Physics, Springer, Berlin, 1979.
\bibitem{newman1} M. E. J. Newman, G. T. Barkema, Monte Carlo Methods in Statistical Physics, Oxford University Press, 2001.
\bibitem{tome} T. Tome, M. J. de Oliveira, Phys. Rev. A 41, (1990) 4251.
\bibitem{kaneyoshi3} T. Kaneyoshi, J. Phys. Condens. Matter 3 (1991) 4497.
\bibitem{hasenbusch} M. Hasenbusch, Phys. Rev. B 84, (2011) 134405.



\end{thebibliography}
\end{document}